\documentstyle[11pt,AATS,epsf]{article}	

\begin{document}	

\title{Is Cold Dark Matter Still a Strong Buy? The Lesson from Galaxy Clusters}

\author{Fabio Governato}
\affil{Osservatorio Astronomico di Brera, Milan, Italy (fabio@merate.mi.astro.it)}
\author{Sebastiano Ghigna}
\affil{Astronomy Department, University of Washington, Seattle, WA, USA}
\author{Ben Moore}
\affil{Physics Department, University of Durham, Durham, UK}




\begin{abstract}
For the last few years the Cold Dark Matter model (ticker: CDM), has
been the dominant theory of structure formation. We briefly review the
recent advancements and predictions of the model in the field of
galaxy clusters.  A new set of very high resolution simulations of
galaxy clusters show that they have (1) density profiles with central
slopes very close to $-1.6$ and (2) abundance of subhalos similar to
the ones observed in real clusters.  These results show a remarkably
small cluster to cluster variation and a weak dependence from the
particular CDM cosmology chosen (LCDM having $\sim 40\%$ less
substructure than SCDM). While still a speculative theory with a high
prediction/evidence ratio, subject to strong challenges from
observational data and competition from other hierarchical theories,
we give CDM a rating of ``market outperform'' and of ``long term
BUY''.
\end{abstract}


\section{Introduction}

Introduced in the early '80s (Peebles 1984; Davis et al.\ 1985), Cold
Dark Matter has rapidly become the dominant model within the
hierarchical clustering framework. Repeat the mantra with us: ``in
this theory primordial density fluctuations collapse and merge
continuously under the effect of gravitational instability to form
more and more massive structures''.

\begin{figure}[htb]	
\plotsetheight{8.5cm}{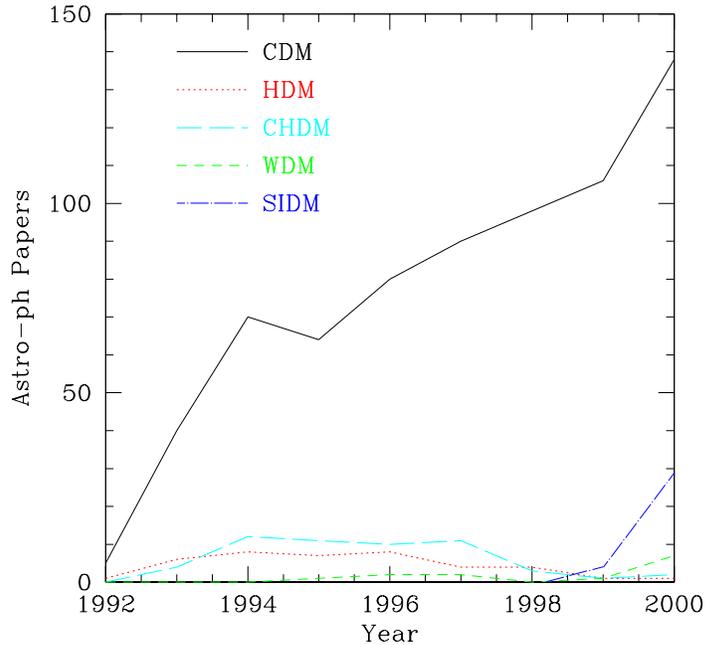}
\caption{The number of  papers on different hierarchical DM  models
submitted to the astro-ph database  from  1994 to 2000.}
\end{figure}

One of the most appealing features of CDM is its ability to give a
solid framework to provide predictions on the astrophysical properties
of cosmological objects, as the number density as a function of mass
and redshift and their clustering properties. All on a range of more
than 10 orders of magnitude in mass and from redshift $\sim 100$ to the
present.

Not bad.

Being the most massive self bound objects in the Universe, galaxy
clusters have received lots of attention, both on the theoretical and
the observational side. Statistical properties of the cluster
population can be obtained using numerical simulations and/or
semi-analytical methods (Governato et al.\ 1999; Jenkins et al.\ 2001;
Sheth, Mo, \& Tormen 2001; but it all started with Press \& Schechter
1974). Cool stuff; however this short review focuses on recent results
obtained using $N$-body simulations on the internal structures of
clusters within the CDM framework.

\section{Is CDM the dominant theory for cosmic  structure formation?}

As with business companies, there are many, often fuzzy, ways to
evaluate the ``dominant position'' of a theory like influential
papers, citations, number of people involved. For CDM a readily
available estimate is the number of papers submitted to the arXiv.org
e-print archive (Greenspan, ehm Ginsparg 1996) in the ``astro-ph''
section. Simple, but fair compared to {pro forma earnings, registered
users or web page hits often used to evaluate some of Nasdaq's (ex)
darlings' performance. Clearly these data show that CDM is the most
widely used cosmological theory for structure formation (see Figure~1)
at least compared to other dark matter models.

The number of papers with the word CDM in the abstract has grown at a
compound rate of about 15\% per year, comparable or higher than the
stock market!  (the well known Dow Jones and S\&P~500 indexes have
long term returns of about 10--15\% per year). CDM has been able to
reinvent itself through the years easily incorporating new
experimental evidence that quickly changed our view of cosmology in
the last decade.  CDM faced its biggest crisis in 1994, due to
mounting criticism against its simplest but very successful product
SCDM, i.e., a critical Universe, 95\% dominated by dark
matter. Problems for the model came from lack of power at large scales
(Efstathiou et al.\ 1990), predicted evolution of galaxy cluster
numbers stronger than observed (Henry et al.\ 1992) and the baryon
fraction in galaxy clusters too low to be reconciled with observations
(White et al.\ 1993).  SCDM had to be recalled from customers and the
following year the number of papers containing CDM in their abstract
declined almost 10\%, while competing models soared, including HDM, a
cosmological model already ruled out in the '80s (White, Frenk, \&
Davis 1983).  Indeed alternative hierarchical models enjoyed then a
moment of success. CHDM introduced a small component of massive
neutrinos (e.g., Ghigna et al.\ 1997) to increase the amount of large
scale power, while other models, like $\tau$CDM or Warm Dark Matter
(Hogan \& Dalcanton 2000) tried to decrease the amount of power at
galactic and subgalactic scales.  However, these days only Self
Interacting Dark Matter (SIDM, see section 4.2) shows a growth rate
higher than CDM, but with only a fraction of its market share.
 
The first robust detection of primordial perturbations in the Cosmic
Microwave Background from the COBE satellite suggested that the CDM
business model was on the right track, although in need of some major
restructuring.  After 1998 and observational evidence for an
accelerating Universe (Perlmutter et al.\ 1999) the new ``Standard''
model became LCDM, a flat Universe with a cosmological constant,
$\Omega_0=0.3 \sim \Omega_{\rm CDM}$ and normalization $\sigma_8 \sim
1$.  Indeed just a few days ago CDM topped analysts expectations after
the findings of optical redshift survey 2dF (Peacock et al.\ 2001) and
the analysis of the full set of BOOMERANG's data (Netterfield et al.\
2001), which strongly supported a LCDM universe with baryon abundance
close to nucleosynthesis predictions.

\begin{figure}[hbt]	
\plotsetheight{9.2cm}{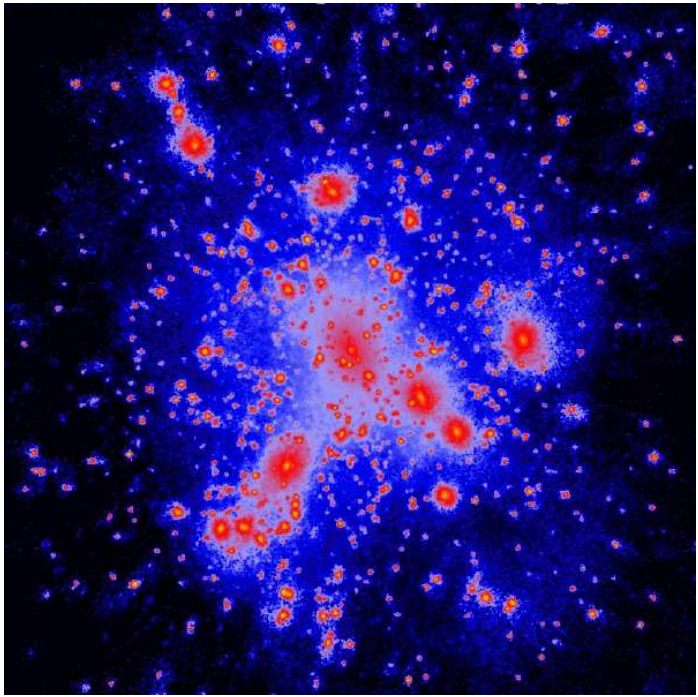}
\caption{The color phase--density plot of a high SCDM resolution
cluster. Box size is twice the virial radius, corresponding to
$1.5h^{-1}\;\rm Mpc$.}
\end{figure}

Interestingly, the fraction of papers submitted to the astro-ph
archive containing the word CDM in the abstract is actually a
diminishing fraction of the total number of papers submitted. It was
10\% in 1994 and only 2\% in the year 2000. Is cosmology going out of
fashion? Are we cosmologists losing market share to planet formation,
AGNs and, perish the thought, funny variable stars? We offer here the
following very speculative (or provocative?)  explanation: the total
number of papers submitted to the preprint database is growing slower
than the total number of world Internet users which doubles every year
or so. This is because scientists have likely been faster to adopt the
Internet than the average population (no AOL or IOL to fight with);
cosmo theorists have been faster than the average astrophysicist
population and their number as users of the database got rapidly close
to 100\%.  It is likely that now virtually all of CDM related papers
are submitted to astro-ph, while other fields in astrophysics are
slower to adopt it as a preferred way to disseminate preprints. The
number of generic astrophysics papers submitted (being low at the
beginning) has a much larger room to grow compared to that of just CDM
papers.

\section{Simulations of Galaxy Clusters}

With the advent of parallel architectures and dedicated hardware
(e.g., GRAPE; Hut \& Makino 1999) and cross testing of $N$-body codes
(e.g., Frenk et al.\ 1999) it has been possible to simulate with
accuracy not only the large scale distribution of galaxy clusters, but
individual objects at a much larger detail. This is intrinsically a
difficult numerical problem, given the large dynamic range across the
cluster and the number of dynamical times (T$_{\rm dy} < 0.01/H_0$) at
its center. Insufficient dynamical range would cause infalling halos
to dissolve in the cluster potential when their central densities
became comparable {\it The effect of increasing the dynamical range in
a simulation is to correctly model the evolution of the densest
structures (e.g., a subhalo core region), allowing them to survive the
tidal forces of the cluster} (Moore, Katz, \& Lake 1996).

A new generation of simulations (Figure~2) has allowed us to test CDM
under a new, interesting aspect: the internal properties of clusters
and galaxies halos, namely the abundance of substructure and the
density profile of the parent dark matter halo. A comparison of their
results (Ghigna et al.\ 1998; Brainerd et al.\ 1998; Tormen, Diaferio,
\& Syer 1998; Klypin et al.\ 1999a; Ghigna et al.\ 2000; Fukushige \&
Makino 2000 among many), suggest that (1) a spatial resolution of less
than a few percent of the virial radius, (2) half a million particles,
(3) several tens of thousands time steps for particles with the largest
acceleration, and (4) a surrounding simulated region of several Mpc
are required to correctly model the tidal field, the subhalo
population and a halo density profile down to a small fraction of its
virial radius.

In this review we will briefly discuss previous results from different
authors and show findings from a new set of very high resolution
simulations (Governato et al., in preparation).  These simulations
explore the cosmic scatter in halo properties (1) within the same
(SCDM) cosmology at a fixed mass, (2) with different power spectra (but
keeping phases fixed), and (3) at different masses in the same (LCDM)
cosmology. The set of  simulations presented here satisfies all the 
requirements of the previous paragraph.

\begin{figure}[hbt]	
\plotsetheight{7.8cm}{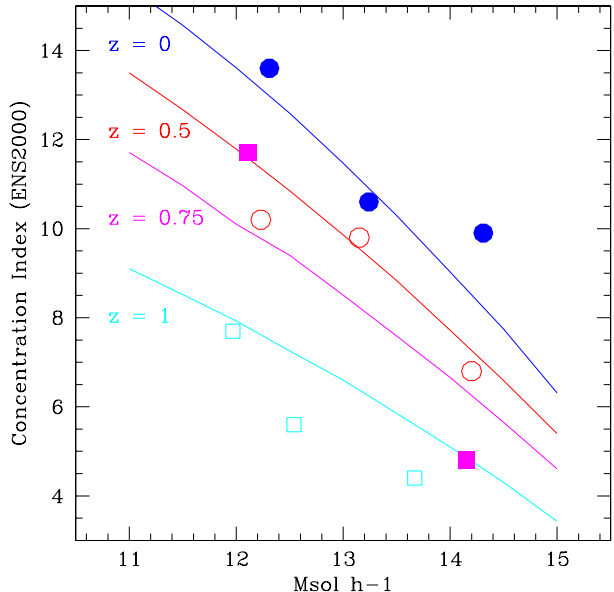}
\caption{The concentration parameter (assuming an NFW profile) for a
set of halos at different redshifts in a LCDM cosmology. Lines are
predictions from ENS. (full dots z=0, empty dots z=0.5, full squares
z=0.75, empty squares z=1). All these halos have at least 30\,000 particles
within the virial radius. Low $z$ ones have half a million or more.}
\end{figure}

\section{The Internal Structure of Galaxy Clusters}

\subsection{Density Profiles}

A significant progress in our understanding of the internal structure
of dark matter halos has been the fundamental finding that halos
formed in CDM cosmogonies follow a universal profile, with a halo
concentration that depends on the amplitude of density fluctuations as
well on the ratio of power at small and large scales (Navarro,
Frenk, \& White 1996, hereafter NFW; Eke, Navarro, \& Steinmetz 2000,
hereafter ENS).

The proposed density profiles are, among others,
\begin{displaymath}
\rho/\rho_{\rm crit} = \frac{\delta_c}{(r/r_{\rm s})(1+r/r_{\rm s})^2}
\end{displaymath}
(NFW) or
\begin{displaymath}
\rho/\rho_{\rm crit} =
\frac{\delta_c}{(r/r_{\rm s})^{1.5}(1+(r/r_{\rm s})^{1.5})}
\end{displaymath}
(Moore et al.\ 1998, 1999a), where $\delta_c$ is a function of the so
called ``concentration parameter'' $c = r_{\rm s}/r_{\rm vir}$,
$r_{\rm vir}$ is the virial radius, and $r_{\rm s}$ is a scale radius.

\begin{figure}[hbt]	
\plotsetheight{9.5cm}{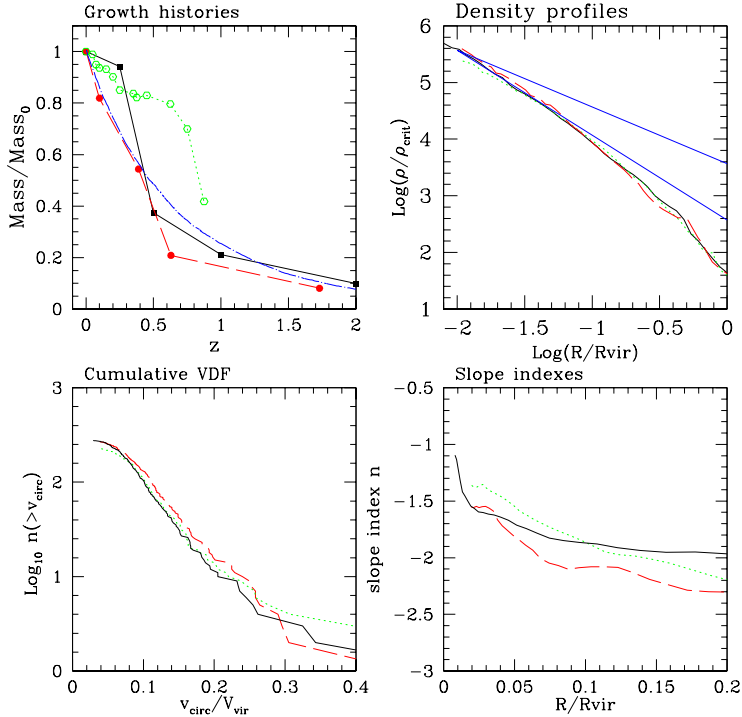}
\caption{Upper left: the growth histories of three Virgo-like ($\sim
4 \times 10^{14}M_\odot$) SCDM clusters vs.\ the average predicted
with the extended Press \& Schechter formalism. Upper right: density
profiles. Straight lines are slopes of $-1$ and $-1.5$
respectively. Lower left: the substructure abundance as a function of
the ratio of circular velocities of subhalos vs.\ the main one (see also
Figure~6). Lower right: effective slope of the density profile (the
continuous line extends to a very small fraction of the virial radius
to show the effect of resolution).}
\label{f:panel4_1}
\end{figure}

\begin{figure}[hbt]	
\plotsetheight{9.5cm}{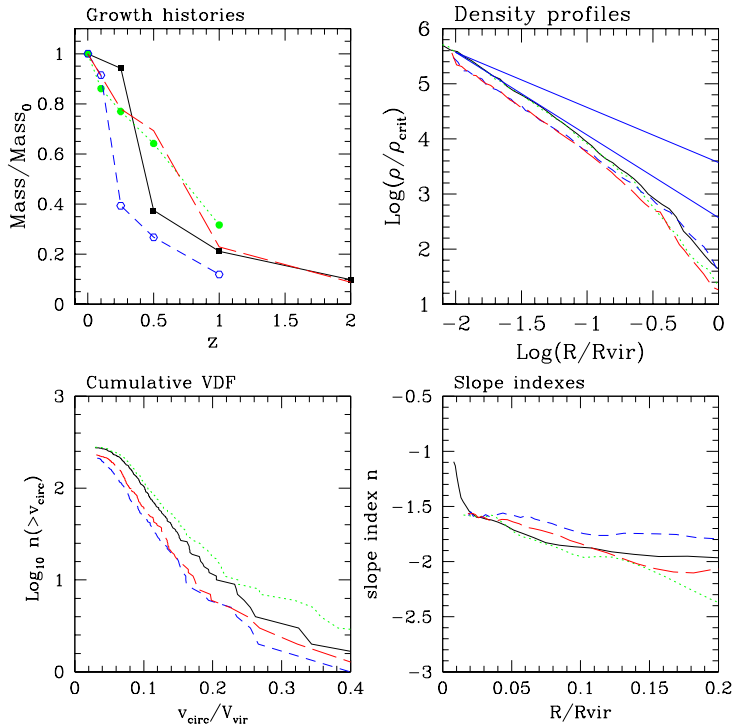}
\caption{Upper left: the growth histories of three Virgo-like clusters
in SCDM (continuous line), LCDM (dashed), OCDM (dotted) \& TCDM (short
dashed) cosmologies.  Models have been cluster normalized.  Upper
right: density profiles. Straight lines are slopes of $-1$ and $-1.5$
respectively. Lower left: the substructure abundance as a function of
the ratio of circular velocities for subhalos vs.\ the main one. Lower
right: effective slope of the density profile.}
\label{f:panel4_2}
\end{figure}

This, after some empirical tuning, allows detailed predictions of the
shape of halo profiles. There is general consensus that in all CDM
variants halo concentrations decrease at higher $z$ at fixed mass
(i.e., at larger $M/M_*(z)$) (see also Bullock et al.\ 2001). However,
parameter space is large and previous works were able to cover a
limited part of it at currently state-of-the-art resolution (you need
to explore different cosmologies, a large range in redshifts and
masses and keep cosmic variance into account before drawing any strong
conclusions). Our new simulations are a step in that direction and
confirm the ENS predictions for the concentrations of halos
(Figure~3).

{\it Note for the profile aficionado: the value of $c$ depends
somewhat on the binning method used to measure the density profiles.
For results in Figure~3 we used a procedure similar to that used in
ENS (Eke, private communication) namely: (1) $\sim 50$ logarithmic
bins between $0<r<r_{\rm vir}$, (2) Poisson weighting, and (3) fitting
between 0.01 (0.02 for $N_{\rm part}<10^5$) and $0.75 r_{\rm vir}$}.

\begin{figure}[hbt]	
\plotsetheight{8.5cm}{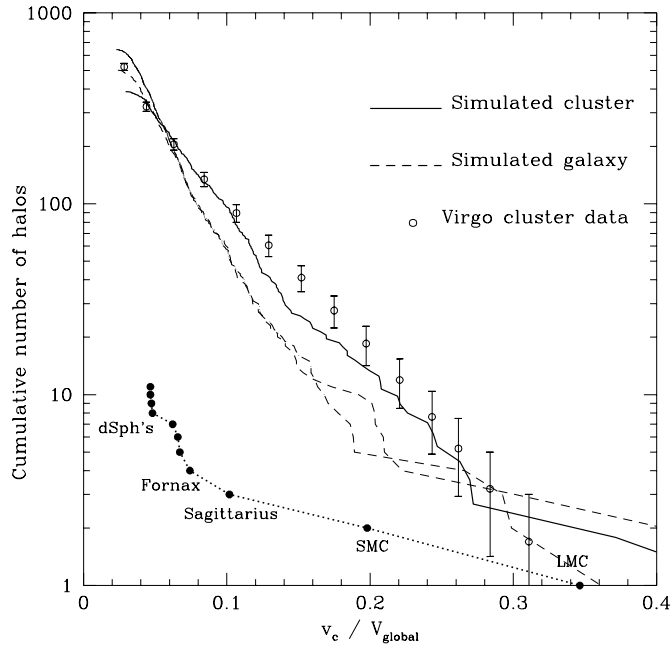}
\caption{The abundance of cosmic substructure within our Milky Way
Galaxy, the Virgo cluster and our simulated models of comparable
masses.  We plot the cumulative numbers of halos as a function of
their circular velocity ($v_{\rm c}=\sqrt{(Gm_{\rm b}/r_{\rm b})}$
normalized to the circular velocity, $V_{\rm global}$ of the parent
halo.  The dotted curve shows the distribution of the satellites
within the Milky Way's halo (Mateo 1998) and the open circles with
Poisson errors is data for the Virgo galaxy cluster (Binggeli, Sandage,
\& Tammann 1985), with galaxy luminosities transformed to circular
velocities using the Tully--Fisher relation.  The second dashed curve
shows data for the galaxy at an earlier epoch, 4 billion years ago.}
\label{f:abundance}
\end{figure}

While interesting, the LCDM model predictions for the concentrations
are difficult to test for cluster sized halos. Low concentrations
(5--10) imply that the change in the density profile slope happens at
relatively large radii of the order of 100$\;$kpc or larger. Eke,
Navarro, \& Frenk (1998) and Carlberg, Yee, \& Ellingson (1997)
reported good agreement between a NFW profile and cluster profiles
from galaxy counts, under the assumption that galaxies trace the
underlying mass distribution (Carlberg et al.\ 1996), a somewhat
reasonable assumption but difficult to test with simulations, as
unwanted numerical effects will tend to underestimate the number of
galaxies in the central part of clusters.

Stronger constraints can be placed at galactic scales, both measuring
the shape of the rotation curves of individual dwarf and LSB galaxies
(Flores \& Primack 1994; Moore 1994), or, perhaps more robustly, the
mass inside the optical radius (ENS) as low resolution rotation curves
have likely been affected by beam smearing (van den Bosch et al.\ 2000).
On the theory side, some claims that LCDM halos were too concentrated
(Navarro \& Steinmetz 2000) have been retracted (ENS) and careful
re-analysis of observational data have been able to set much weaker
constraints on the theory (van den Bosch \& Swaters 2001).

A tad confusing, isn't it?

In our opinion the crucial point is the central slope of the density
profile in CDM dark matter halos. This is an issue far from being
settled. The two proposed profiles have substantially different
profile slopes in the inner part of the halo, converging to $-1$ and
$-1.5$.  While current observations seem to be able to accommodate
slopes as steep as $-1$ within a few per cent of the virial radius,
halos profiles as steep as $-1.5$ or more, as shown by Moore et al.\
(1999a) and recently by Fukushige \& Makino (2001) would prove rather
difficult to support. Obviously firmer predictions have to be made to
use high resolution rotation curves and mass profiles from weak
lensing and X-ray observations (e.g., Lombardi et al.\ 2000) to
establish whether there is a strong LCDM crisis.

With the aim of settling the issue of the central slope of the density
profile in clusters we have performed a number of high resolution runs
of Virgo-sized clusters (${\rm a~few} \times 10^{14}M_\odot$). Three
halos were taken from a SCDM cosmology to address the issue of cosmic
scatter.  Another Virgo-sized halo was run in four different
cosmologies (LCDM, SCDM, TCDM, \& OCDM), but keeping the same phases.
  
{\it In all cases the slope of the density profile within $0.01 <
r_{\rm vir} < 0.1 $ is very close to $-1.6$, significantly steeper
than the central slope advocated by NFW} (Figures 4 and 5).

\subsection{Substructure in Galaxy Clusters}

Increasing the dynamical range of numerical simulations showed another
major success of the model: as gravitational clustering creates
(statistically) small halos first, some of them get gradually
subsumed into larger halos.  While early works assumed that these
subhalos would have been destroyed (White \& Rees 1978), high res
simulations (e.g., Ghigna et al.\ 2000 and references therein) showed that
they survive within the virialized regions of the parent halos and the
abundance of dark substructures predicted by SCDM agrees well with the
observed abundance of galaxies inside clusters
(Figure~\ref{f:abundance}). {\it Our set of simulations reveals
surprisingly little scatter between different realizations
(Figure~\ref{f:panel4_1}) (contrary to results obtained at much lower
resolution) and between different variants of the CDM model
(Figure~\ref{f:panel4_2})}. At a circular velocity of $\sim$ 200
Km/sec ($v_{circ}/V_{vir}$=0.2) the LCDM cluster has about 40\% less
halos than the SCDM one.

Due to the almost power-law shape of the CDM power spectrum $P(k)$ and
the long survival times of subhalos, there is also little dependence
on the parent halo mass, i.e., once rescaled to the circular velocity
of the main halo, the properties of subhalos of galactic and cluster
halos look pretty much the same (again Figure~\ref{f:abundance}).
While these results are a major success for CDM at cluster scales,
galactic dark subhalos are predicted far in excess of the observed
population of observed galactic satellites, by almost two orders of
magnitude (Moore et al.\ 1999b; Klypin et al.\ 1999b).

Several solutions to this puzzle have been suggested within the CDM
framework (e.g., the association of dark subhalos with High Velocity
Clouds or the effect of an ionizing UV background and SN feedback,
e.g., Moore 2001) but the question is still open.  Other solutions are
being explored.  Self Interacting (or Collisional) Dark Matter which
IPOed just last year (Spergel \& Steinhardt 2000) and WDM (Bode,
Ostriker, \& Turok 2001) are indeed interesting alternatives (or
rather modifications) to LCDM. However, numerical tests of SIDM on the
cluster mass scale have given negative (Yoshida et al.\ 2000) or mixed
results (Moore et al.\ 2000) or require the DM cross section to be a
function of velocity, unlikely in the Newtonian regime (Firmani et
al.\ 2000).

\subsection{Orbits of Galaxies}

Knowing the shape and evolution of orbits of galaxies in clusters is
crucial for dynamical estimates of cluster masses (e.g., van der Marel
et al.\ 2000).  As tidal stripping is very efficient at decreasing a
subhalo mass after the first pericentric passage, subsequent evolution
of the subhalo population appears to be very slow, with a time scale
likely larger than a Hubble time. Ghigna et al.\ (1998) showed clearly
that orbital properties of subhalos do not differ significantly from
those of the underlying DM distribution. Surviving subhalos are on
almost radial orbits with a typical pericenter/apocenter ratio of 1:5.
As subhalos orbit inside the dense background (comprising $\sim85$\%
of the mass of a cluster) they slowly lose orbital energy and sink to
the center. However this process is not very efficient.  $N$-body
simulations have been combined with semi-analytical models to give
insight on the dynamical evolution of the halos identifiable with the
hosts of luminous Lyman Break Galaxies (the most massive halos at $z
\sim 3$) and the progenitors of present day giant ellipticals
(Governato et al.\ 2001).  Orbital shapes of massive halos that fell
into the cluster at high $z$ did not show any statistical difference
from the global halo population, showing that {\it orbital decay and
evolution of surviving galaxies in clusters is negligible over a
Hubble time even for those massive halos that were able to survive as
separate entities in the early phases of the cluster formation}.

Colpi, Mayer, \& Governato (2000) have proposed a theoretical model
for dynamical friction and a fitting formula which keeps orbit
eccentricity and the retarding effect of tidal stripping into account:
\begin{displaymath}
\tau_{\rm DF} = 1.2 \frac{J_{\rm cir}r_{\rm cir}}%
{[GM_{\rm sat}/{\rm{e}}]\ln(M_{\rm halo}/M_{\rm sat})} \varepsilon^{0.4} ,
\end{displaymath}
where $J_{\rm cir}$ and $r_{\rm cir}$ are, respectively, the initial
orbital angular momentum and the radius of the circular orbit with the
same energy of the actual orbit and $\varepsilon$ is the orbit
circularity.  The agreement between the semi-analytical approach and
$N$-body simulations is rather remarkable.

\section{Discussion}

While CDM faces considerable challenges from observational data and
competing theories we believe it will still be the reference model for
years to come. Recent observational results give support to its
business model (but careful investors should perhaps remember the old
saying ``buy low \& sell high''...).  The CDM picture gives a coherent
frame consistent with large scale structure constraints where galaxies
in clusters form in the right numbers and range of masses, almost
independently of cosmology.  Their sizes and masses are governed by
simple and reasonably understood processes like tidal stripping and
dynamical friction. Mass attached to individual galaxies is of the
order of 15\%, a predictions that will be tested by weak lensing
measurements in galaxy clusters (e.g., Natarajan et al.\ 1998). There
is mounting evidence that the inner slope of the dark matter profile
for clusters in CDM models is close to $-1.6$, with a small cluster to
cluster scatter and weak dependence on the cosmological model.

While CDM seems well positioned, daring colleagues and students in
search of market-beating returns should also invest their time and
efforts in competing theories which, while riskier and (more)
speculative, will offer insight on the physical processes linked to
the formation and evolution of galaxy clusters. Space for improvements
is getting tight, as constraints from large scale structures improve,
and deviations from the currently preferred LCDM model will likely
involve galactic and subgalactic scales, with hopefully interesting
implications on our understanding of galaxy formation, star formation
and feedback on the Intra Cluster/Galaxy Medium, especially at high
redshift.

\section{Disclosure and Disclaimer (Conforming to SEC Regulations)}

FG, SG and BM own shares of CDM since the early '90s and have started
some rather speculative investments in WDM (FG) and Collisional DM
(BM)$\ldots$ so some conflict of interests here. Oh well, sue us.
This document contains ``forward looking statements''. These
statements are subject to risks and uncertainties and are based on the
beliefs and assumptions of the writers based on information currently
available. Most important, always remember: past performance is no
guarantee of future success!

\acknowledgments 

The authors thank their colleagues Tom Quinn and Joachim Stadel for
allowing them to show results from ongoing projects.  Simulations were
completed at the ARSC (Fairbanks, AK) and CINECA (Bologna, Italy)
supercomputing centers. FG acknowledges generous support from the
organizers of this conference. Finally, Hawaii Volcanoes National
Park is a way cool place. Go visit it.

\end{document}